\documentclass[11pt,twoside]{article}
\usepackage{asp2010}
\usepackage{natbib}

\aspvolume{471} 
\aspvoltitle{Origins of the Expanding Universe: 1912-1932}
\aspcpryear{2013} 
\aspvolauthor{Michael J. Way and Deidre Hunter, eds.}

\begin{document}

\title{Dismantling Hubble's Legacy?}
\author{Michael J. Way$^{a,b}$
\affil{$^a$NASA/Goddard Institute for Space Studies, New York, NY, 10029 USA}
\affil{$^b$Department of Astronomy, Uppsala University, Uppsala, Sweden}}

\begin{abstract}
Edwin Hubble is famous for a number of discoveries that are well known to
amateur and professional astronomers, students and the general public.
The origins of these discoveries are examined and it is
demonstrated that, in each case, a great deal of supporting evidence
was already in place.  In some cases the discoveries
had either already been made, or competing versions were
not adopted for complex scientific and sociological reasons.
\end{abstract}

\section{Introduction}

Edwin Hubble is considered one of the titans of early 20th century
observational cosmology. He is credited in most
textbooks\footnote{See \citet[][p. 98]{Smith2009JHA....40...71S}.} and
the popular literature for a series of important discoveries
made between 1920 and 1930:

\begin{itemize}
\item The confirmation of the Island Universe hypothesis
\item The classification of extragalactic nebulae
\item The discovery of a linear relationship between distance
and velocity for extragalactic nebulae, providing the first evidence
for the expanding universe
\item The brightness profile of galaxies
\end{itemize}

The discoveries above are well-known to the astronomy community
and most astronomers would associate them solely with Edwin Hubble; yet
this is a gross oversimplification. Astronomers and historians
are beginning to revise that standard story and bring a more
nuanced version to the public's attention. This paper is adding
to this burgeoning reappraisal.\footnote{Some examples
include \cite{nussbaumer2009discovering, KraghSmith2003HisSc..41..141K,bartusiak2010day}.}

As a (small) counter-narrative,
William \citet[][p. 411]{Hoytnational1980biographical},
in his biographical memoir of V.~M. Slipher exclaims that ``[Slipher]
probably made more fundamental discoveries than any other observational
astronomer of the twentieth century."\footnote{Also see
\cite{Hall1970S&T....39...84H}.}  Clearly some historians in the 1970s
and 1980s thought that Slipher made more fundamental discoveries than
Hubble.\footnote{See contributions in this book by John Peacock,
Joseph S. Tenn, Robert Smith, Laird Thompson and Kevin
Schindler for more on Slipher's discoveries.} Yet how can that be true
given all we \emph{know} today? In this paper we re-examine
Hubble's discoveries in some detail in order to see if they
are better understood in a broader context.
Given the focus on V.~M. Slipher at this conference we will also 
explicitly discuss his contributions in two of the cases above.

\section{Discovery of the Island Universe}

The hypothesis of Island Universes has a long history going back
at least to the 18th Century with contributions by:
\cite{Swedenborg1734,Wright1750,Kant1755}\footnote{Kant actually
cited the work of Thomas \cite{Wright1750}.}, and
\cite{Lambert1761cbud.book.....L}.\footnote{See contribution by Ayala
in this book.}
William \cite{Herschel1785RSPT...75..213H} initially believed
that the spiral nebulae were external to the Milky Way, but
later changed his mind. Knut \citet[][Chapter 1]{Lundmark1927UGC..........1L}
does an excellent job of explaining the origins of the Island Universe
that I do not believe has been much bettered by time.\footnote{The
essence of the hypothesis, in an early 20th century context, was
that the universe was populated by many Milky Way galaxies known then
as spiral nebulae. This was opposed to the belief that the universe
consisted of a single Milky Way object with satellites such
as spiral nebulae, globular clusters and Magellanic cloud-like objects.}

To get from philosophical speculation to modern quantification one must
fast-forward to the late 19th and early 20th century to find a large
number of investigations of objects termed ``Nebulae" with the new
art of photography and ever larger telescopes.\footnote{See
\cite{Gingerich1987JRASC..81..113G} for more on this early period.}
For example, \cite{Huggins1864RSPT..154..437H} were deeply interested in the
spectra of nebulae, while astronomers such as
Isaac \cite{Roberts1903AN....160..337R} 
built photographic catalogs.\footnote{Only later was Roberts' catalog
compiled and completed by his wife Mrs. Isaac Roberts.} Using the catalogs
of nebulae like that of Roberts' and
others \citet[][pg.869]{Lundmark1925MNRAS..85..865L}
claimed that nearly 1200 spiral nebulae proper motions had been
measured at that time.  We now know this claim was incorrect -- most
likely it was an incorrect assessment of observational
errors.\footnote{Only in 2012 was the proper motion of M31 possibly
measured by \cite{Sohn2012ApJ...753....7S} using optical observations.}
The field was clearly in its infancy, but progress on distance estimates
to objects like globular clusters and spiral nebulae was moving rapidly forward.

Table \ref{table1} lists all of the main distance estimates to
spiral nebulae (known to this author) from the late 1800s until 1930
when standard candles began to be found in spiral nebulae.

\begin{table}
\caption{Early distance estimates to Spiral Nebulae}\label{table1}
\vspace{0.2cm}
\begin{small}
\begin{tabular}{|l|c|r|l|} \hline
Reference & Object & Distance\tablenotemark{a}    & Method \\
\hline
\cite{Herschel1786RSPT...76..457H} & M31      & $<$17,200 \tablenotemark{b} & color/magnitude\\
\cite{nichol1850architecture}      &``cluster"& 154,800\tablenotemark{c} & magnitude comparison\\
--                                 &          & 302,505 & -- \\
\cite{Clerk1890Book}               & M31      & 564? & nova of 1885 \\
\cite{Clerk1903Book}               & M31      & $<$1000 &  Size \\
\cite{Bohlin1907AN....176..205B}   & M31      & 19   & parallax\\
\cite{Very1911AN....189..441V}     & M31      & 4,000 & diameters \\
\cite{Very1911AN....189..441V}     & M31      & 1,600 & S Andromedae \\
\cite{Wolf1912AN....190..229W}     & M31\tablenotemark{d}& 32,000 & diameters \\
\cite{Curtis1915PASP...27..214C}   & spirals  & 10,000 & astrometry/radial velocity\\
\cite{Pease1916PNAS....2..517P}   & NGC 4594  & 25,000 & astrometry/radial velocity\\
\cite{Curtis1917PASP...29..206C}   & M31      & 20,000,000  & novae \\
--				   & --       & 100,000  & novae\tablenotemark{e}\\
\cite{Shapley1917PASP...29R.213S}  & M31      & 1,000,000 & ``bright stars" \\
\cite{vanMaanen1918PASP...30..307V}& M31      & 250       & parallax \\
\cite{Lundmark1919AN....209..369L} & M31      & 650,000   & novae \\
\cite{Curtis1920JRASC..14..317C}   & misc     & 4,000,000 & novae \\
--				   & misc     & 1,000,000 & novae \\
--				   & misc     &   500,000 & novae \\
\cite{Lundmark1921PASP...33..324L} & M33      & 1,000,000   & ``bright stars" \\
\cite{Luplau1922AN....215..285L}   & M31      & 326,000   & novae\tablenotemark{f} \\
\cite{Opik1922ApJ....55..406O}     & M31      & 1,500,000 & luminosity/mass \\
\cite{Hubble1922ApJ....56..400H}   & M33      & 100,000   & ``stars" \\
\cite{Shapley1923BHarO.796S...1S}  & NGC 6822 & 1,000,000 & diameters/``bright stars"\\
\cite{Hubble1925PA.....33..252H}   & M31/33   & 930,000   & Cepheids \\
\cite{Hubble1925ApJ....62..409H}   & NGC 6822 & 700,000   & Cepheids,``bright-stars"\\
\cite{Lundmark1925MNRAS..85..865L} & M31,M87  & 1,400,000 & novae \\
--				   & --       & 8,000,000 & novae \\
\cite{Lundmark1925MNRAS..85..865L} & M104     & 56,000,000 & \cite{Opik1922ApJ....55..406O} method \\
\cite{Hubble1926ApJ....63..236H}   & M33      & 850,000   & Cepheids,Blue-Giants \\
\cite{Hubble1929ApJ....69..103H}   & M31      & 900,000   & Cepheids,novae\\
M31 value (Dec. 2012)\tablenotemark{g}   & M31      & 2,588,440 & 19 Methods\\ 
\hline
\end{tabular}
\end{small}
\tablenotetext{a}{Units of light years}
\tablenotetext{b}{Herschel stated on page 262 that ``...I believe to be an
indication that its distance in this coloured part does not exceed 2000 times
the distance of Sirius." Using the modern value of the distance to Sirius of 8.6
light years yields an upper limit of 17,200. Note that no parallax measurement
to a star had yet been achieved.}
\tablenotetext{c}{Estimated the maximum distance a \emph{cluster} could be resolved
using Herschel's telescope to be either 18,000 or 35,175 times the distance to
Sirius (p. 51). The modern distance to Sirius was used as above.}
\tablenotetext{d}{Believed the Milky Way to be $\sim$1000 light years in
diameter, so this number is well outside the Milky Way by his own estimate.  Wolf also
measured distances to a number of other Spiral Nebulae, e.g. M33 (86,000 light
years), M81 (170,000), M101 (270,000), M51 (310,000) -- all well outside
the Milky Way.}
\tablenotetext{e}{Curtis claimed that the distances to the novae
in Andromeda were 100 times farther away than the galactic ones. The galactic
novae were estimated to be 1000 light years away.}
\tablenotetext{f}{They used two methods, one like that of Lundmark (1919)
using comparable brightnesses of novae in M31 and our Galaxy, and the other involving
the distances between novae in M31 and our Galaxy.}
\tablenotetext{g}{From the NASA Extragalactic Database. The value presented here
is an average (as of December 2012) from 19 different methods and 133 data points: 
http://ned.ipac.caltech.edu/cgi-bin/nDistance?name=MESSIER+031}
\end{table}

The ability to estimate \emph{accurate} distances of objects beyond the
reach of parallax only came into being with the publishing of
the period--luminosity relationship for Cepheid Variable stars by
Henrietta Leavitt \& Edward Pickering (1912)
and its later calibration by Ejnar Hertzsprung (1913);
Henry Norris Russell (1913)\footnote{Using 13 Cepheids} and later
Harlow \cite{Shapley1918CMWCI.151....1S}\footnote{Using 11 Cepheids of
Hertzsprung's original 13.} all utilizing the
Lewis \cite{Boss1910pgcs.book.....B}
catalog of proper motions.\footnote{Although Shapley's distances were strongly
contested by a number of people including \cite{Curtis1921national.bulletin}.}
Before Cepheids were discovered in spiral nebulae there were a number of
attempts to use novae as standard candles to measure the distances
to spiral nebulae \citep{Curtis1917PASP...29..206C,Shapley1917PASP...29R.213S,
Lundmark1919AN....209..369L,Luplau1922AN....215..285L}.  For example,
Heber \cite{Curtis1917PASP...29..206C} calculated an average distance to
the spiral nebulae of 20,000,000 light years in one case and found them to be
around 100 times as distant as the galactic novae in another.
Lundmark obtained a distance to Andromeda of 650,000 light years.
\cite{Shapley1918CMWCI.151....1S} attempted to compare the
``brightest stars" in our own galaxy to that of Andromeda and stated:
\begin{quote}
... the minimum distance of the Andromeda
Nebula must be of order a million light years. At that remote distance
the diameter of this largest of spirals would be about 50,000 light years a
value that now appears most probable as a minimum for our galactic
system.\footnote{This is ironic given his later disavowals of his larger distance
estimates and those related to novae in
spiral nebulae \citep{Shapley1919ApJ....49..311S}.}
\end{quote}

Initially the novae studies in Andromeda were difficult to
reconcile with a supernovae observed in Andromeda $\sim$35 years previously
\citep{Krueger1885AN....112..245K,Hartwig1885AN....112..285H,
Hartwig1885AN....112..355H,deVaucouleurs1985ApJ...295..287D}\footnote{This 
supernova was later denoted `S Andromedae'. At that time supernovae were
unknown so they were easily confused with normal novae.}, but given the multiple
observations of fainter novae observed in spiral nebulae Curtis (and later
others) was persuaded to drop S Andromedae and
Z Centauri\footnote{Z Centauri (in NGC 5253) was another {\it bright} nova
observed by \cite{Pickering1895HarCi...4....1P}.} as anomalies.

Unfortunately for the novae derived distance measurements,
other observations at that time called their accuracy into question.
In the mid--1910s \cite{Curtis1915PASP...27..105C}
and then \cite{Lampland1916PA.....24..667L}
detected rotation in spiral nebulae, but the former did not believe his
own detection while the latter's results were not
influential \citep[][p. 31]{Smith1982expanding}.  However, additional
observations of this sort by \cite{vanMaanen1916CMWCI.118....1V}
in Messier 101 (M101) and later in Messier 33 (M33) and other nebulae
\citep{vanMaanen1923CMWCI.260....1V} along with the support
of James \cite{Jeans1917Obs....40...60J} convinced many
astronomers like \cite{Shapley1919PASP...31..261S}
that novae derived distances to spiral nebulae were impossible to
reconcile without superluminal speeds of spiral nebulae
rotation.\footnote{This was also the time when Shapley came up
with his 300,000 light year diameter Milky Way galaxy, much
larger ($>$ 30 times) than any other estimate at that time.}

However, in spite of the confusing novae observations and (incorrectly) observed
rotation of spirals the evidence continued to mount that the spiral nebulae
were indeed very distant objects.
The first evidence of this was a fascinating paper by
\cite{Opik1922ApJ....55..406O} where ``an expression is derived for the
absolute distance in terms of the linear speed $v_{o}$ at an angular distance
$\rho$ from the center, the apparent luminosity $i$, and $E$, the energy
radiated per unit mass." He calculated a distance of 1.5 million light
years to M31.  One year later \cite{Shapley1923BHarO.796S...1S} used
diameters of galaxies and the brightness of super-giant stars in
NGC 6822 to state:

\begin{quote}
The above considerations all indicate that the distance of N.G.C. 6822 is
of the order of a million light years. It appears to be a great star cloud
that is at least three or four times as far away as the most distant of
known globular clusters and probably quite beyond the limits of the galactic
system.
\end{quote}

This quote is particularly interesting in light of Shapley's long standing
opposition to the Island Universe hypothesis and his super-galaxy model
\citep{Shapley1921national.bulletin}, but he still would not let
go of his super-galaxy model just yet.

The issue was effectively settled by two papers from Hubble in 
1925 in which he derived distances from Cepheid variables found in M31 and M33
\citep{Hubble1925PA.....33..252H} of 930,000
light years\footnote{The original Cepheid that Hubble discovered in M31 now has
a modern ephemeris and light curve published by \cite{Templeton2011PASP..123.1374T}.}
and in NGC 6822 \citep{Hubble1925ApJ....62..409H} of 700,000 light years.
Note that there were no citations to previous distance estimates in the former
paper and only a reference to \cite{Shapley1918CMWCI.151....1S}
for his calibration of the Cepheid variable light curves in the latter.

Still, there was some confusion about van Maanen's spiral nebulae observations
among his contemporaries including Knut Lundmark. 
\cite{Lundmark1922PASP...34..108L} was at first dismissive of van Maanen's
measurements because of a preponderance of conflicting data. However, by 1922
he had changed his mind after having measured some of van Maanen's plates
himself \citep{Lundmark1927UGC..........1L}.\footnote{See page 17 where he
states that [in 1922?] ``When remeasuring Messier 33 during my stay at Mount
Wilson the situation seemed to be rather hopeless for the followers of
the island--universe theory."} Upon re-measuring the motions in M33
a couple of years later, Lundmark concluded that van Maanen's measurements
were flawed \citep{Lundmark1926ApJ....63...67L,Lundmark1927UGC..........1L}.
However, it would not be until 1935 that \cite{vanMaanen1935ApJ....81..336V}
would nearly admit that his measurements of the rotation of spiral nebulae were
false. In the same issue of \emph{The Astrophysical Journal}
\cite{Hubble1935ApJ....81..334H} published
his own measurements showing any measured rotation to be within the measurement
errors. Clearly Hubble wanted to make sure that the persistent observations
of van Maanen were dismissed by publishing his own measurements given
his (now) elevated status as one of the more highly respected astronomers of
his day.\footnote{\citet[][Chapter 11]{Christianson1996} explains
what happened in more detail.} Thus it took nearly a decade after Hubble's 1925
paper for van Maanen's measurements to be completely disposed of and
the Island Universe theory to be confirmed. Of course many astronomers felt
the matter had been settled all the way back
in 1926.\footnote{As mentioned above
\citep{Lundmark1926ApJ....63...67L} but also \cite{Luyten1926NH.....26..386L}:
``It is now universally accepted that the spiral nebula are millions of
light years distant."}

Still, as mentioned by
Robert \citet[][p. 114]{Smith2008JHA....39...91S} ``...what was
missing to settle the dispute on the spirals was a method of calculating their
distances that a great majority of astronomers could agree was accurate."
Clearly \cite{Hubble1925PA.....33..252H} provided that method with
his observations of Cepheids in spirals, but a great many
people before him made his observations possible. As
\citet[][p. 74]{Smith2009JHA....40...71S} points out ``... it is appropriate
to view Hubble as confirming rather than discovering the extragalactic
nature of spirals. But, following the dictum of John Herschel that he
who proves discovers, Hubble was given the credit."
Many important contributions to this story have been forgotten
and most textbooks in astronomy today, if they discuss the
``Island Universe" confirmation at all, bestow 100\% of the credit on
Hubble with scant attention to the earlier observations that clearly
supported his measurements.

At that time the use of Cepheids as standard candles was considered a
very reliable method of estimating distance. However, by the mid-1950s
it had become clear that Hubble's distances measurements contained significant
systematic errors. Recalibration of the Cepheids by
Walter \cite{Baade1956PASP...68....5B} later helped to show that Andromeda was twice
as far away, and was actually larger than our own Milky Way.
This would have serious implications for the Big Bang theory
in the 1930s and 1940s.

\subsection{Slipher's Contribution to the Island Universe story}

The anniversary date for this conference was intended to overlap
with the published date of Slipher's first observation of a doppler
shift in a spiral nebula (Andromeda) on
17 September 1912 \citep{Slipher1913LowOB...2...56S}. He obtained
an astounding value of --384 km s$^{-1}$. This was surprising because
it was nearly an order-of-magnitude higher than any other measured
doppler shift in the heavens at that time. By 1917 Slipher had observed
25 spiral nebulae, the largest having a redshift of 1100 km s$^{-1}$
\citep{Slipher1917PAPhS..56..403S}. As we have seen above the
debate on whether spiral nebulae were Island Universes went on until
Hubble discovered Cepheids in Andromeda and other spiral nebulae.
Given the 25 spiral nebulae with radial velocities discovered 
by Slipher in 1917 (21 of which were redshifts) why didn't the astronomical
community realize these objects could not be bound to the Milky Way
and must be Island Universes? In fact a number of people did reach
this conclusion including \cite{Campbell1917Sci....45..513C} and
Hertzsprung (see Robert Smith's chapter in this book).  Still,
it took several years for Slipher to convince the community that what he was
observing was real. As well the reticence to push this interpretation
by Slipher himself was related to his modest
personality.\footnote{See Section \ref{sec:mythic} and
Robert Smith's chapter in this book.} In reality what
made it difficult (and would make it difficult for \emph{all} proponents
of an Island Universe theory) were the observations of internal
motion in Spiral Nebulae by Adrian van Maanen that began with
his first publication on the subject in July of 1916
\citep{vanMaanen1916PNAS....2..386V}.\footnote{See David DeVorkin's contribution
to this proceedings and the first chapter in \cite{Smith1982expanding}.}
Had it not been for the erroneous observations of van Maanen it is
likely that Slipher's observations would have provided
strong support for the Island Universe theory.

\section{Classification of Extra-galactic Nebulae}

The first systematic classification of nebulae was probably attempted by
\cite{Herschel1786RSPT...76..457H} in his paper titled ``Catalogue of One
Thousand New Nebulae and Clusters of Stars." Therein he described eight different
classes of objects:
\begin{enumerate}
\setlength{\itemsep}{1pt}
  \setlength{\parskip}{0pt}
  \setlength{\parsep}{0pt}
\item Bright nebulae [93 examples]
\item Faint nebulae [402]
\item Very faint nebulae [376]
\item Planetary nebulae [29]
\item Very large nebulae [24]
\item Very compressed and rich clusters of stars [19]
\item Pretty much compressed clusters of large or small stars [17]
\item Coarsely scattered clusters of stars [40]
\end{enumerate}
His descriptions of the nebulae were extremely detailed using
terms with single letter abbreviations, for example: B. Bright, S. Small, v. very,
e. extremely, R. Round, M. in the middle, l. a little, g. gradually, 
r. resolvable, m. milky. These were used in combinations, one of his
own examples being vgmbM: (v)ery (g)radually (m)uch (b)righter in the (M)iddle.

Later Lord \cite{Rosse1850PhilTran...XXV..499R} gave the term spiral to some of
Herschel's nebulae by using his new 1.8m telescope ``Leviathon of
Parsonstown," but he described
it first via a drawing of M51 presented to the 15th meeting of the British
Association for the Advancement of
Science \citep{Rosse1845BAAS...36R,Hoskin1982JHA....13...97H,Dewhirst1991JHA....22..257D}.

\begin{table}
\centering
\caption{Early Classification schemes for Extragalactic--Nebulae}\label{table2}
\vspace{0.2cm}
\begin{tabular}{|l|l|} \hline
Reference & Notes \\
\hline
\cite{Herschel1786RSPT...76..457H} & ``first comprehensive scheme?"\\
\cite{Rosse1850PhilTran...XXV..499R}& terminology ``Spirals" used\\
\cite{Wolf1908PAIKH...3..109W}     & ``widely cited scheme" \\
\cite{Bailey1908AnHar..60..199B}   & --\\
\cite{Pahlen1911AN....188..249P}   & --\\
\cite{Bigourdan1914CRAS......1949B}& --\\
\cite{Shaw1915Helwan15}            & --\\
\cite{Curtis.lick1918studies}      & ``bars"    \\
\cite{Curtis1919..........98C}     & --\\
\cite{Jeans1919CosmogStellDyn}     & Theoretical \\
\cite{Reynolds1920MNRAS..80..746R} & Classification of spirals like \cite{Hubble1922ApJ....56..162H}\\
\cite{Hubble1922ApJ....56..162H}   & Preliminary scheme \\
\cite{Lundmark1926ArMAF..19B...8L} & Preliminary scheme \\
\cite{Hubble1926ApJ....64..321H}   & More complete scheme \\
\cite{Lundmark1927UGC..........1L} & Full scheme \\
\cite{Shapley1927BHarO.849....1S}  & --\\
\cite{Jeans1928astronomy}          & Tuning-fork diagram suggestion\\
\citep{Hubble1936realm}            & Tuning fork diagram added to create complete scheme\\
\hline
\end{tabular}
\end{table}

The classification scheme of \cite{Herschel1786RSPT...76..457H} (with
later modifications by son John Herschel) was considered unwieldy and
complicated, but was probably the only one referred to consistently until new
schemes in the early 20th century such as that
of \cite{Wolf1908PAIKH...3..109W}.\footnote{The
classification scheme of \cite{Wolf1908PAIKH...3..109W} is mentioned in
a number of articles, but without a proper citation.
The UGC catalog of \citet[][p. 452]{Nilson1973ugcg.book.....N} cites the year
1909: ``The well-known classification system for nebulae, the Wolf
code a-w with 23 standard objects, was presented for the first
time in 1909." It was actually 1908.} Wolf's classification scheme
worked for all types of nebulae.
He not only lists specific examples of each but also includes a table
of images. He labeled them with letters `a--w' (there is no letter `j', but
rather an `h' and `h$_{o}$'). Another interesting scheme was developed
by \cite{Bigourdan1914CRAS......1949B}.

Today \cite{Hubble1926ApJ....64..321H} is generally given credit for coming
up with the first usable classification scheme of ``Galaxies", or as they
came to be known ``Extra-galactic nebulae". In fact the Table in his 1926 paper
is titled ``Classification of Nebulae" which included both ``Galactic nebulae"
and ``Extra-galactic nebulae" which is an extension of his earlier work
\citep{Hubble1922ApJ....56..162H}. Using his 1922 work as a basis, he had
tried to build his Extra-galactic nebulae classification scheme in-line with
the nebular evolutionary model of \cite{Jeans1919CosmogStellDyn}.

One of the great strengths of the \cite{Hubble1926ApJ....64..321H} paper was
his formula that described the spiral divisions:
\begin{equation}\label{eq:1}
m_{t}=C-5log(d)
\end{equation}
where $m_{t}$$=$total magnitude, $d=$diameter of the nebulae, $C=$Constant describing
each object in his sequence (1--3): Sa(1),Sb(2),Sc(3),SBa(1),SBb(2),SBc(3)
where 1$=$Early, 2$=$Intermediate, 3$=$Late. Hence as a nebulae aged from
``Early" to ``Late" the diameter and luminosity would change accordingly
and (again) in-line with the theoretical work of \cite{Jeans1919CosmogStellDyn}.
However, other models at that time \citep{Lindblad1927MNRAS..87..420L}
contradicted some aspects of Jean's evolutionary sequence
\citep{Jeans1919CosmogStellDyn}.\footnote{In
particular \cite{Lindblad1927MNRAS..87..420L}
states, ``We do not assume a general development from less flattened to more
flattened system of higher angular speed of rotation [like that of Jeans]."
Lindblad also believed that the centers of the the nebulae were simply unresolved
faint stars -- contrary to Jeans.}

David Block \& Ken Freeman (2008)
appear to most recently describe how Hubble's entire ``Extra-galactic
nebulae" classification scheme was remarkably similar to one developed by
John \cite{Reynolds1920MNRAS..80..746R}. In fact they present clear evidence
that Hubble (at this time) was aware of
the \cite{Reynolds1920MNRAS..80..746R} paper via an unpublished memo written to
Reynolds, which they reproduce in their book.
However, \cite{Hubble1922ApJ....56..162H,Hubble1926ApJ....64..321H} does not
cite \cite{Reynolds1920MNRAS..80..746R} in these papers, although he does give
credit to \cite{Curtis.lick1918studies} for the recognition of bars in
spiral nebulae.


A year after he introduced his 1926 classification scheme,
\cite{Hubble1927Obs....50..276H} mentioned a paper from earlier
in 1927 in which \cite{Reynolds1927Obs....50..185R} criticizes
Hubble's published classification scheme of 1926. Hubble again omits mention of
\cite{Reynolds1920MNRAS..80..746R} while Reynolds not only mentions Hubble's
work, but an even earlier classification scheme by \cite{Shaw1915Helwan15}.

What makes Hubble's omission of \cite{Reynolds1920MNRAS..80..746R} particularly
troubling is that he accused Lundmark of plagarism not only in personal
correspondence\footnote{He asked Lundmark to explain himself and threatened to
publish his accusation \citep[][p. 103]{Holmberg1999reaching}.},
but also on page 3 of his 53-page classification scheme
article \citep{Hubble1926ApJ....64..321H}:\footnote{It was published a few months
after \citep{Lundmark1926ArMAF..19B...8L}.}

\begin{quote}
Meanwhile K. Lundmark, who was present at the Cambridge meeting and has
since been appointed a member of the Commission, has recently published
(Arkiv f\"{o}r Matematik, Astronomi och Fysik, Band 19B, No.8, 1926) a
classification, which, except for nomenclature, is practically identical
with that submitted by me. Dr. Lundmark makes no acknowledgments or references
to the discussions of the Commission other than those for the use of the
term `galactic'.
\end{quote}

This is rather remarkable because in no paper published by Hubble between
1920 and 1930 is the classification scheme of \cite{Reynolds1920MNRAS..80..746R}
mentioned.  \cite{Lundmark1927UGC..........1L} strongly rebutted Hubble on
page 24 of his 127-page paper titled ``Studies of Anagalactic
Nebulae."\footnote{See Appendix \ref{appendixlundmarkreply}
for Lundmark's full reply.} The latter denotes
the classification scheme of Wolf in one of the columns\footnote{No citation is
provided, but he must be referring to \cite{Wolf1908PAIKH...3..109W}.} next
to his own, but also mentions other work that preceded his (page 23):
\begin{quote}
Classifications of nebulae based on photographic material have been
made by Bailey, Curtis, Mrs. Isaac Roberts, Max Wolf, Hubble and others.
\end{quote}

We do not have the space here to delve deeply into the personalities
of Hubble or Lundmark, yet we may get some feeling for what their
contemporaries felt about them and how they felt about their
contemporaries via the limited notes placed in papers and in
their personal correspondence. Some of the latter can be found in
\cite{Smith1982expanding}, while some specific examples in
the case of Hubble are described
in \citet[][Chapter 11]{Christianson1996}.\footnote{A particular
quote from Walter Adams, then the director at Mt. Wilson, is
worth repeating in reference to the conflict between Hubble and
van Maanen over the distances to the spirals: ``This is not the
first case in which Hubble has seriously injured himself in the opinion of
scientific men by the intemperate and intolerant way in which he has expressed
himself." This was in reference to the conflict between
Hubble and van Maanen that Adams had to negotiate as director.}

On the other hand Lundmark has been called enigmatic by \cite{Smith1982expanding},
but at least some of his contemporaries appreciated his general attitude. Take
this quote from Ludwik \cite{Silberstein1925MNRAS..85..285S}:
\begin{quote}
... I should like to express my deep gratitude to Dr. Lundmark for having
devoted so much attention to the discussion of this problem from a perfectly
impartial attitude.
\end{quote}
This was in reply to a paper by \cite{Lundmark1924MNRAS..84..747L} that critized
\cite{Silberstein1924MNRAS..84..363S}
for his use of globular clusters (GCs) to determine the curvature radius of the
Universe. Lundmark felt that GCs were not distant enough.\footnote{It may be
amusing to note that six years later \cite{Lundmark1930PA.....38...26L} asked
whether the GCs and Extragalactic (he used the word Anagalactic)
nebulae were related.} \cite{Holmberg1999reaching} in his Chapter titled
``Lundmark and the Lund Observatory" also paints a picture of a complex
character, but whose bitterness towards some of his Swedish colleagues appeared
to surface later in his career when his scientific productivity was waning.

Finally, it is clear that Lundmark had been thinking of a classification scheme
for nebulae at least since 1922 \citep{Teerikorpi1989JHA....20..165T}
and even discussed a simplified ``class of objects" in
\cite{Lundmark1925MNRAS..85..865L}.  Lundmark's scheme for ``Anagalactic
nebulae"\footnote{His name for ``Extra-galactic nebulae."}
was broken into 4 groups (see page 22 of \cite{Lundmark1927UGC..........1L}):
1.) Anomalous nebulae (Aa), 2.) Globular, elliptical, elongated, ovate or
lenticular nebulae (Ae), 3.) Magellanic (``irregular") nebulae (Am),
4.) Spiral nebulae (As), where the degree of condensation toward the center
was his main criteria. On the other hand \cite{Hubble1926ApJ....64..321H}
separated his ellipticals by eccentricity. Spirals were separated based on form
and degree of arm development.  In fact the Lundmark and Hubble schemes were
{\it not} considered the same by their
contemporaries.\footnote{See \citet[][p. 152]{Smith1982expanding}.}

A year later \cite{Shapley1927BHarO.849....1S} joined in the classification
attempts with a model that incorporated aspects of the work of both Lundmark and
Hubble, but his model was not adopted. Contrary to Hubble, Shapley carefully
cited his predecessors including \cite{Bailey1908AnHar..60..199B,
Reynolds1920MNRAS..80..746R,Wolf1908PAIKH...3..109W,
Hubble1922ApJ....56..162H,Lundmark1926ArMAF..19B...8L,Hubble1926ApJ....64..321H}.

Hubble's classification scheme is also noted for its later tuning fork
design to separate the barred spirals from
non--barred ones \citep{Hubble1936realm}.
\cite{Block.Freeman2004ASSL..319...15B} have pointed out that Hubble was not the
first to describe the tuning fork diagram -- that was originally proposed
by \cite{Jeans1928astronomy}.

It is generally acknowledged that Hubble's classification scheme became
standard because it had an evolutionary component and mathematical
description (see Equation \ref{eq:1}) that previous schemes did not.
But as should be clear from Table \ref{table2} there were a great
many classification schemes leading up to that of Hubble's which
surely influenced him. Is it troubling that Hubble does not readily cite two of
the most important and influential schemes (before his own was published) of
\cite{Reynolds1920MNRAS..80..746R} and \cite{Wolf1908PAIKH...3..109W} and yet
accuses a contemporary (Lundmark) of plagarism on the basis of
scant evidence? This lack of citation by Hubble will be further discussed
in Section \ref{sec:literature}.

\section{Discovery of the ``Hubble Constant"}

A great deal has been written in recent years on the topic of the
discovery of the expanding universe 
\citep{nussbaumer2009discovering,Shaviv2011arXiv1107.0442S,
KraghSmith2003HisSc..41..141K,Smith1982expanding}.
A number of accusations have been levelled against
Hubble \citep{Block2011arXiv1106.3928B}, some of which have been
discredited \citep{Livio2011Natur.479..171L}.
Several chapters in this book contain discussions on the discovery
of the expanding universe
(see chapters by Cormac O'Raifeartaigh, Ari Belenkiy, Harry Nussbaumer,
John Peacock, and Robert Smith).  For that reason there is no
need to go into a great deal of detail here, but suffice
it to say that this ``discovery" is even more complicated than the other
stories described above. In Table \ref{table3} one can see a steady progression
of three related measures: 1.) the solar motion with respect to the nebulae, 
2.) the radius of curvature of the universe and 3.) the linear relation of
velocity and distance for the spiral nebulae that lent
support to an expanding universe model over a static one
in the first half of the 20th century. Note that papers that
do not explicitly discuss observational data are not included in the table. 
Some further details on selected publications from Table \ref{table3} are
worth mentioning in detail:

\begin{table}
\footnotesize
\caption{Early estimates of solar motion, curvature radius and H$_{o}$ via
spiral nebulae}\label{table3}
\begin{center}
\begin{tabular}{|l|c|l|} \hline
Reference & Date & Notes \\
\hline
\cite{Truman1916PA.....24..111T}$^{a}$      &1915/12/30 & RA=20h, Dec=-20$\deg$, V=--670km s$^{-1}$ (14 spirals) \\
\cite{Young-Harper1916JRASC..10..134Y}$^{a}$&1916/02/01 & RA=20h24m, Dec=-12$\deg$10$\arcmin$, V=--598$\pm$234km s$^{-1}$\\
                                            &           & (15 spirals + MC)\\
\cite{Paddock1916PASP...28..109P}$^{b}$     &1916/05/00 & V=--295$\pm$202 km s$^{-1}$, K=+248$\pm$88 km s$^{-1}$\\
                                            &           & Using \cite{Young-Harper1916JRASC..10..134Y} data.\\
\cite{Wirtz1916AN....203..293W}             &1916/08/00 & Various values of RA, Dec, V, but no K term\\
\cite{Wirtz1917AN....204...23W}             &1916/12/06 & Various values of RA, Dec, V, but no K term\\
\cite{Slipher1917PAPhS..56..403S}           &1917/04/13 & RA=22h, Dec=-22$\deg$, V=--700 km s$^{-1}$\\
\cite{deSitter1917MNRAS..78....3D}          &1917/07/00 & First estimates of R for Models A and B\\
\cite{Wirtz1918AN....206..109W}             &1917/12/00 & K=+656 km s$^{-1}$, V=--830 km s$^{-1}$\\
\cite{Shapley1919ApJ....50..107S}           &1918/11/00 & Magnitude vs. Velocity\\
\cite{Lundmark1920KSVH...60....1L}          &1920/01/26 & K=+587 km s$^{-1}$ (29 spirals, page 75)\\
\cite{Wirtz1922AN....215..349W}	            &1921/10/00 & K=+656 km s$^{-1}$, V=--820 km s$^{-1}$\\
\cite{Friedmann1922ZPhy...10..377F}         &1922/05/29 & Set M=5$\times10^{21}M_{\sun}$, $\lambda$=0 and found\\
                                            &           & ``world period''=10$^{10}$ years\\
\cite{Wirtz1924AN....222...21W}             &1924/03/00 & Distance vs. velocity: $v$(km)=2200--1200$\times$log($Dm$)$^{c}$\\
\cite{Silberstein1924MNRAS..84..363S}       &1924/03/00 & Distance vs. velocity Relation and calculation of\\
                                            &           & curvature radius (R)\\
\cite{Silberstein1924Natur.113..350S}       &1924/03/08 & Calculates R for 11 GC$^{d}$\\
\cite{Silberstein1924Natur.113..602S}       &1924/04/26 & First distance vs. velocity plot for \emph{GC and LMC/SMC}\\
\cite{Lundmark1924MNRAS..84..747L}          &1924/06/00 & First distance vs. velocity plot for \emph{spiral nebulae},\\
                                            &           & K$_{spirals}$= +800 km s$^{-1}$\\
\cite{Silberstein1924Natur.113..818S}       &1924/06/07 & Same method and data as before with new estimate of R\\
\cite{Silberstein1924Natur.114..347S}       &1924/10/00 & \\
\cite{Silberstein1924theory}                &1924/Late  & Distance vs. velocity plot\\
\cite{Silberstein1925MNRAS..85..285S}       &1925/01/00 & R=7.2$\times10^{12}$ A.U. and updated plot of distance\\
                                            &           & vs. velocity for GC+MC+M33\\
\cite{Stromberg1925ApJ....61..353S}         &1925/06/00 & Estimates of K, but no relation found for\\
                                            &           & distance vs. velocity\\
\cite{Lundmark1925MNRAS..85..865L}          &1925/06/00 & Re-defines: K=k+lr+mr$^{2}$, First time for a\\
                                            &           & variable K--term\\
\cite{Dose1927AN....229..157D}              &1926/11/00 & K=+765$\pm$111 km s$^{-1}$ for spirals (no variable K--term)\\
\cite{Lemaitre1927ASSB...47...49L}          &1927/04/00 & Discovers that K is linearly dependent on distance\\
\cite{Robertson1928PhilMag}                 &1928/00/00 & Expects a correlation between velocity and distance \\ 
\cite{Hubble1929PNAS...15..168H}            &1929/01/17 & Distance vs. Velocity plot using Cepheid distances\\
                                            &           & yields a linear fit\\
\cite{deSitter1930BAN.....5..157D}          &1930/05/26 & Using observational data calculates R and estimates\\
                                            &           & slope of velocity vs. distance \\
\cite{Hubble1931ApJ....74...43H}            &1931/03/00 & Updated list of distances and velocities yields\\
                                            &           & 558 km s$^{-1}$ Mpc$^{-1}$ \\
\cite{Oort1931BAN.....6..155O}              &1931/11/30 & Finds H$_{o}$=290 km s$^{-1}$ Mpc$^{-1}$ after finding \\
                                            &           & some distance inaccuracies\\
\hline
\end{tabular}
\end{center}
\tablenotetext{a}{Solar motion relative to Spiral Nebulae, MC=Magellanic Clouds}
\tablenotetext{b}{First appearance of K correction}
\tablenotetext{c}{Dm = Distance via diameter}
\tablenotetext{d}{GC = Globular Clusters}
\end{table}

\begin{itemize}

\item O.H. \cite{Truman1916PA.....24..111T}: was the first to measure the solar motion
relative to the spiral nebulae like that of \cite{campbell1913stellar} and
\cite{Airy1860MemRAS..28..143A}:\newline
V = X cos$\alpha$ cos$\delta$ + Y sin$\alpha$ cos$\delta$ +
Z sin$\delta$ \footnote{$\alpha$=Right Ascenson, $\delta$=Declination,
X,Y,Z = velocity components of our sun through space.}

\item George \cite{Paddock1916PASP...28..109P}: realized there may be a non-random
spiral nebulae recessional component (denoted as K). He didn't believe it was
real, but others quickly thought otherwise. This paper contains
the first appearance of the ``K
correction"\footnote{Not to be confused with the modern usage of ``K correction"
which refers to resetting the colors of a galaxy to the rest frame.}
in the formula for solar motion:\newline
V = X cos$\alpha$ cos$\delta$ + Y sin$\alpha$ cos$\delta$ + Z sin$\delta$ + K.\footnote{K is the average recessional velocity of spiral nebulae
observed in km s$^{-1}$.}

\item Willem \cite{deSitter1917MNRAS..78....3D}: attempted to measure the radius of curvature
of the universe (R) for Einstein's model A and his own model B in a number of ways.
In model A he made an estimate of the mass and total volume of the universe
to obtain R$\leq$5$\times$10$^{13}$
Astronomical Units (A.U.)\footnote{\citet[][p. 25]{deSitter1917MNRAS..78....3D}}
(790$\times10^{6}$ light years). 
In model B he made a number of estimates (all found in Section 6 of his paper),
but perhaps the most relevant are:
	\begin{enumerate}
	\item ``If we accept the existence of a number of galactic systems whose
average mutual distances are of the order 10$^{10}$ all we can say is that $\pi$R
must be several times 10$^{10}$ or roughly $R>10^{11}$[A.U.]."
(1.6$\times10^{6}$ light years)
	\item ``For the lesser Magellanic cloud Hertzsprung found the distance
r$>$6$\times$10$^{9}$. The radial velocity is about +150 km sec$^{-1}$. This gives
$R>2\times10^{11}$[A.U.]." (3$\times10^{6}$ light years)
	\item By averaging the velocities of 3 spiral nebulae (+600 km s$^{-1}$) and
their distances (326,000 light years) he obtained $R=3\times10^{11}$A.U.
(4.7$\times10^{6}$ light years)
	\end{enumerate}

\item \cite{Shapley1919ApJ....50..107S}: ``The speed of spiral nebulae is
dependent to some extent upon apparent brightness, {\it indicating a relation
of speed to distance} or, possibly, to mass." [our emphasis]

\item Alexander \cite{Friedmann1922ZPhy...10..377F}: derived the first non-static solutions
in addition to the solutions of \cite{Einstein1917SPAW.......142E} (model A) and
\cite{deSitter1917MNRAS..78....3D} (model B). He estimated the age of the
universe, but we do not know where he obtained his numbers from. He set the mass
of the universe M=5$\times10^{21}M_{\sun}$, set $\lambda$=0 and found a
``world period"=10$^{10}$ years. However see Ari Belenkiy's chapter in this
book. Belenkiy believes this number should have been written 10$^{9}$ years.

\item \cite{Silberstein1924MNRAS..84..363S}: estimated a distance vs. velocity
relation for 7 GC, LMC and SMC\footnote{GC=Globular Clusters, LMC=Large
Magellanic Could, SMC=Small Magellanic Cloud)},
R$=$94$\times$10$^{6}$ light years (R=6$\times$10$^{12}$ A.U.).
His estimate was later criticized by \cite{Lundmark1924MNRAS..84..747L} (among others)
for only using 7 of 16 known GC. \cite{Lundmark1924MNRAS..84..747L} states: ``... there
is no good reason for selecting such an arbitrary limit for excluding objects
{\it which do not give a rather constant value of R.}"
\cite{Lundmark1924MNRAS..84..747L} also complained that the low value of K (+31
for GCs vs. +800 km s$^{-1}$ for spiral nebulae) ``suggests that the former are
comparatively near as compared with the latter", implying that they were
inappropriate for calculating R. [Published March, 1924]

\item \cite{Silberstein1924Natur.113..350S}: for 11 GC R$<$94$\times$10$^{6}$
light years (R$<$6$\times$10$^{12}$ A.U.) [Published March 8, 1924]

\item \cite{Silberstein1924Natur.113..602S}: first distance vs. velocity plot
with 2 fits for 11 \emph{Globular Clusters} including the LMC and SMC.
(one fit only used 8 GC).  R$<$158$\times$10$^{6}$ light years
(R$<$1$\times$10$^{13}$ A.U.). [Published April 26, 1924]

\item \cite{Silberstein1924Natur.113..818S}: used the same objects as in
\cite{Silberstein1924Natur.113..602S} and found R=110--126$\times$10$^{6}$
light years (R=7--8$\times$10$^{12}$ a.u.).  [Published June 7, 1924]

\item \cite{Lundmark1924MNRAS..84..747L}: first distance vs. velocity plot
for \emph{spiral nebulae}. Included separate plots for GC+LMC+SMC,
and a variety of stellar types, but without any lines fit to any samples since he did
not feel it was warranted. He used the novae distances to Andromeda
and then used spiral nebulae diameters (assuming constant nebular diameters and
luminosities) for the other spirals in his sample in comparison with
Andromeda -- his x-axis units were in `Distances of Andromeda Nebula.'
He also calculated K=+800 km s$^{-1}$ (for spirals), R=19.7$\times$10$^{12}$ A.U.
(3$\times$10$^{8}$ light years) for GC (but with a very large dispersion
of 26.1$\times$10$^{12}$ A.U.)\footnote{Lundmark stated: ``As the dispersion in R
is 26.1$\times$10$^{12}$ km and thus considerably higher than what could be
expected from the dispersions in V and r, it does not seem that the curvature
of space-time, at least for the present, can be determined with any accuracy
by using the displacements in the spectra of globular clusters."}
and 2.4--6.6$\times$10$^{12}$ A.U.
(3.8$\times$10$^{7}$--1$\times$10$^{8}$ light years) for spirals.\footnote{Note: Lundmark
quoted R in units of kilometers, but it is clear in comparisons with
Silberstein's papers and with his own distance estimates to GCs and spiral
nebulae that he must have meant to write A.U. rather than km.}
\cite{Duerbeck2001mkt..book..231D} fit Lundmark's data in the plot for
spirals using Lundmark's smaller value of the distance to Andromeda of 0.2 Mpc
to yield a slope with the origin through zero of 90 km s$^{-1}$ Mpc$^{-1}$.
If one uses an Andromeda distance of 0.5 Mpc (derived by Lundmark using novae)
one obtains 36 km s$^{-1}$ Mpc$^{-1}$. Also see Peacock's contribution in this volume.
[Published June, 1924]

\item \cite{Silberstein1924Natur.114..347S}: R=7.2$\times$10$^{12}$ A.U.
(1.1$\times$10$^{8}$ light years). [Published September 6, 1924]

\item \cite{Silberstein1924theory}: distance vs. velocity plot for 11 GC, plus
the LMC, SMC {\it and} M33.  Values for R were nearly the same as
in \cite{Silberstein1924MNRAS..84..363S}. [Published late 1924]

\item \cite{Silberstein1925MNRAS..85..285S}: newly updated distance vs. velocity
plot using 18 GC, the LMC, SMC, and M33.
Obtained R=7.2$\times10^{12}$ A.U. (1.1$\times$10$^{8}$ light years).
[Published January 1925]

\item \cite{Stromberg1925ApJ....61..353S}: after an extensive investigation into
spiral nebulae (and separately GCs) he considered correlations between
distance vs. velocity
and positions on the sky (cos $\lambda$) vs. velocity. He ended his article by
stating: ``In conclusion we may say that we have found no sufficient reason to
believe that there exists any dependence of radial motion upon distance from the
sun.  The only dependence fairly well established is one that is a function of
position in the sky." He plotted \emph{two} equivalent correlations for the globular
clusters on the same plot with wildly different slopes and stated: ``It is
significant, however, that the regression-line for the clusters does not
go through the origin as expected from the theory."
[Published June 1925]

\item \cite{Lundmark1925MNRAS..85..865L}: initially believed the K-term was a
constant for spirals, but decided it was given by $K=k+lr+mr^{2}$. Here
k,l,m are constants, and the r is relative distance via the apparent diameter.
Solving with 44 velocities gave k=513,l=10.365,m=0.047.

\item Georges \cite{Lemaitre1927ASSB...47...49L}: derived a non--static solution
to Einstein's equations and coupled it to observations to reveal
a linear distance vs. redshift relation with a slope of 670 or
575 km s$^{-1}$ Mpc$^{-1}$ (depending on how the data is grouped).
Radial velocities were from \cite{Stromberg1925ApJ....61..353S},
distances from apparent magnitudes given in \cite{Hubble1926ApJ....64..321H}
that were taken from \cite{Hopmann1921AN....214..425H} and
\cite{Holetschek1907AWS....20.....H}.

\item Howard Percy \cite{Robertson1928PhilMag}: ``... we should nevertheless
expect a correlation
v$\approxeq$cl/R between assigned velocity v, distance l, and radius of the observable
world R." [equation 17] Using the data of \cite{Hubble1926ApJ....64..321H} for distances
and Slipher \citep{Eddington1923mathematical} for velocities he obtained R=2$\times10^{27}$cm
(1.3$\times10^{14}$ A.U., 2.1$\times10^{9}$ light years).
Hilmar Duerbeck \& Waltraut Seitter (1999) have estimated
his distance vs. velocity slope as 460 km s$^{-1}$ Mpc$^{-1}$. Robertson also says
that a similar relation to that of equation 17 was deduced by
\cite{Weyl1923PZ...24..230W}.

\item \cite{Hubble1929PNAS...15..168H}: used Cepheids and bright stars for
distances and spiral nebulae Doppler shifts mostly from
Slipher \citep{Eddington1923mathematical}. He found a linear relation
between distance and velocity using the data available
(grouping them two ways) and an updated solar motion equation:
V=Xcos$\alpha$cos$\delta$+Ysin$\alpha$cos$\delta$+Zsin$\delta$+kr, where
the old K is now a function linearly dependent upon distance (K=kr).
He quoted a slope of $\sim$465 $\pm$50 km s$^{-1}$ Mpc$^{-1}$ for 24
objects, and $\sim$513 $\pm$60 km s$^{-1}$ Mpc$^{-1}$ for 9 groups.
He stated: ``The outstanding feature, however, is the possibility that
the velocity-distance relation may represent the de Sitter effect, and hence
that numerical data may be introduced into discussions of the general curvature
of space."

\item \cite{deSitter1930BAN.....5..157D}: used observational data from nebulae
and calculated the slope of the velocity vs. distance linear fit: V/cr=0.5$\times10^{-27}$
c.g.s (V/r$\sim$450 km s$^{-1}$ Mpc$^{-1}$) which for model A yields
R$_{A}=2.3\times10^{27}$cm = 1.5$\times10^{14}$A.U. and for model
B yields R$_{B}=2\times10^{27}$cm = 1.3$\times10^{14}$A.U.

\end{itemize}

Perhaps the most notable name that readers will find missing from the table and individual
descriptions is Arthur Eddington. Eddington took part in this project in important ways
that did not include actual ``discoveries":

\begin{enumerate}
\item He participated in a number of important discussions with most of the authors listed
in Table \ref{table3}.\footnote{See \cite{Smith1982expanding}}

\item He was responsible for the re-publication and translation of Lema\^{i}tre's
1927 paper in \emph{Monthly Notices of the Royal Astronomical Society}
\citep{Lemaitre1931MNRAS..91..483L}. He initially
brought Lema\^{i}tre's work to the attention of the world in his
May 1930 paper \citep{Eddington1930MNRAS..90..668E}.

\item He published the final list of Slipher's radial velocities \citep{Eddington1923mathematical}.

\item By some he is even considered to be the {\it transition} figure who
triggered the major paradigm change from the ``static or stationary" model of
the universe to an ``evolving geometry"
\citep{Ellis1990Innovation}.\footnote{See page 98, Table 6.1.}

\end{enumerate}

Overall we find that Lema\^{i}tre was the first to seek and find a linear
relation between distance and velocity \emph{in the context of an expanding universe},
but that a number of other actors (e.g. Carl Wirtz, Ludwik Silberstein,
Knut Lundmark, Edwin Hubble, Willem de Sitter) were looking for a relation
that fit into the context of de Sitter's
Model B world with its {\it spurious} radial velocities. This is discussed
in a number of other papers in this book (see contributions
by Harry Nussbaumer, Cormac O'Raifeartaigh, and Ari Belenkiy).

\subsection{Slipher's Contribution to the Expanding Universe Story}

Slipher's radial velocities played a critical role in \emph{all} of the
publications listed in Table \ref{table3} above. Lets look at Slipher's
data in several of the most important papers in this table.

\cite{Lundmark1925MNRAS..85..865L} used Slipher's radial velocity data of spirals
to look for a relation between distance and velocity. While he did not cite
Slipher's work, he did state on page 866, ``Mainly on account of the enthusiastic
and skilful work of V.~M. Slipher we have now knowledge of 44 radial velocities
of spiral nebulae." Why wouldn't Lundmark cite Slipher's work containing the
44 radial velocities he used? In fact Slipher never published his final list
of radial velocities, the final list was found in Arthur Eddington's book of
1923 \citep{Eddington1923mathematical}.\footnote{See the contribution
by Robert Smith in this book, and Slipher's table of velocities in
Eddington (1923) reproduced in Ari Belenkiy's contribution.}

\cite{Lemaitre1927ASSB...47...49L} also used the radial velocities of Slipher,
but Slipher's name did not appear in this paper. Rather he cited the work
of \cite{Stromberg1925ApJ....61..353S} as his source.
\cite{Stromberg1925ApJ....61..353S} listed 56 velocities obtained from
Slipher (it included some globular clusters in addition to spiral nebulae),
but stated ``Slipher's determinations are given without references\dots ." 
Str\"{o}mberg otherwise praised Slipher in his Introduction stating
``..but through the perseverance of Professor V.~M. Slipher, a fairly
large number of such velocities has been derived." Perhaps Lemaitre could
be forgiven as he was mainly a theorist, but it's troubling that he didn't
take the time to cite the original sources of his data.

\cite{Hubble1929PNAS...15..168H} used Slipher's radial velocities for
20 out of 24 objects \footnote{The numbers come from Peacock's chapter
in this book.} listed in his famous Figure 1 showing a
``Velocity--Distance Relation among Extra-Galactic Nebulae". Hubble gave
no attribution to Slipher in this paper, only stating that ``Radial velocities
of 46 extra-galactic nebulae are now available\dots ." However, he did give credit
to a few people, mentioning that two of the distances listed in his Table 1 were
those of Shapley (he gave no citation), three velocities were those of Humason,
and with the exception of three measured by himself, the rest of the visual
magnitudes listed were ``Holetschek's visual magnitude as corrected by Hopmann."

\subsection{Hubble Finds his Expanding Universe?}

It is commonly believed that Hubble not only discovered an expanding universe,
but that he was also looking for it. The former is credible, but the latter
is not. Thus far historians have unearthed no evidence that Hubble was searching
for the clues to an expanding universe when he published his 1929
paper \citep{Hubble1929PNAS...15..168H}. Given the timing of events
it is difficult to reconcile. There were only a few people with
knowledge of a non-static solution to Einstein's equations in 1928-29:
\begin{enumerate}
\item Alexander Friedman: passed away in 1925.
\item Yuri Krutkov \& Paul Eherenfest who worked to get Friedman's papers
published and negotiated with Einstein over their validity (see Belenkiy in this
book).
\item Georges Lema\^{i}tre: his 1927 paper was published in French in an obscure
Belgian journal. He sent his paper to at least Einstein, de Sitter and Eddington.
\item Einstein: discussed Lema\^{i}tre's paper with him at the 1927 Solvay conference,
\footnote{See Belenkiy's chapter in this book.} but told him he did not believe
in his solution. For the first time Lema\^{i}tre also learned from Einstein of
the Friedman 1922 and 1924 papers.
\item De Sitter: it is not clear that he ever read Lema\^{i}tre's 1927 paper
prior to 1930.
\item Eddington: appears to have forgotten about Lema\^{i}tre's 1927 paper
until he was reminded of it in
early 1930.\footnote{See \cite{nussbaumer2009discovering}, Chapter 11.}
\end{enumerate}

It was not until May 1930 when the papers of \cite{Eddington1930MNRAS..90..668E}
and \cite{deSitter1930BAN.....5..157D} were published that the rest of the world
became aware of the non-static solutions of Lema\^{i}tre and later the earlier
solutions of Friedman.

\section{Brightness Profile of Galaxies}

\cite{Reynolds1913MNRAS..74..132R} was perhaps the first to attempt
the measurement of the light profile of the Andromeda Galaxy, but 
only across the bulge, not out to the spiral arms. His careful measurements
yielded:
\begin{equation} \label{eq:2}
Luminosity = \frac{Constant}{(x+I)^{2}}
\end{equation}

\noindent where $x=$ distance from the center of the nucleus/bulge along the major
axis (out to a diameter of 7$\arcmin$).

Seven years later \cite{Reynolds1927Obs....50..185R} went after more nuclei
in a number of spiral nebulae (M65, M99, M100, M94, and M64) but with
mixed success at applying Eq. \ref{eq:2}.

Three years later \cite{Hubble1930CMWCI.398....1H} describes a ``Distribution
of Luminosity in Elliptical Nebulae." In this particular case Hubble {\it does}
cite the \cite{Reynolds1913MNRAS..74..132R,Reynolds1927Obs....50..185R} papers,
but perhaps since Reynolds was focused on the bulges of spiral nebulae
rather than Ellipticals Hubble didn't feel the need to ignore his competitor.
Regardless, Hubble later generalized the relation for Ellipticals as:

\begin{equation} \label{eq:3}
I=\frac{I_{o}}{\left( \frac{r}{a} + I \right) ^{2}}
\end{equation}


\citet[][p. 133]{Hubble1930CMWCI.398....1H} also gives credit where credit is due:
\begin{quote}
The pioneer investigations along this line are due to J. H. Reynolds,
who, in 1913 found that the luminosity along the major axis of M 3I,
out to 7$\arcmin$ from the nucleus, could be represented by the formula
L=Constant/(x+I)$^{2}$.
\end{quote}

Here we have a (unique?) case where Hubble has properly cited and praised
his predecessor.  This relation is now referred to as the Hubble Luminosity
Profile, but perhaps it would be more properly named the Reynolds--Hubble
Luminosity Profile.

In the end it is perhaps not so relevant as there are a number of other far more
popular profiles in use today including
the \cite{deVaucouleurs1948AnAp...11..247D}, \cite{King1962AJ.....67..471K}
and \cite{Sersic1963BAAA....6...41S} profiles.

\section{Combing Through the Literature}\label{sec:literature}

It may be possible to better quantify Hubble's unwillingness to cite
his predecessors by examining the literature more closely.
Most bibliographies in the 1920s were
not compiled at the back of each article as it is done today. A work
was cited by the person's name and the citation would be contained
in a footnote at the bottom of the same page. For this reason the
SAO/NASA Astrophysical Data Service (ADS) does not have complete bibliographies
for the papers of the period of interest. This author took all of the scientific
publications of Hubble from 1920 through 1930 and attempted to put
together a pseudo-bibliography for each paper. Thankfully Hubble
did not write that many papers in comparison with
Lundmark (see Appendices \ref{appendixHubble} and \ref{appendixLundmark}). A pseudo-bibliography
here means that if someone's name was mentioned in the context of a previous
publication but no bibliographic information was included it was included
as a citation. Full citations are also included in the counting.

Knut Lundmark was chosen as a comparative figure to Hubble as he was
probably considered by his peers to be a figure of equal stature during the
1920s \footnote{This was more so in the beginning of the 1920s, less so
in the latter after Hubble's many discoveries.}. Lundmark
wrote many more papers than Hubble, but ADS does not have a
complete set of the papers he wrote during this period of time.
A handwritten book was obtained \footnote{The author is unknown, hence
there is no citation for it.}
from the Uppsala University Library that contained a listing of
all of Lundmark's papers in his career (even it is not complete, but
more so than ADS).  Some of the papers missing from ADS found in this book are
included in Appendix \ref{appendixLundmark}.  Papers were not included that
were considered popular science.\footnote{Lundmark wrote at least 35 articles
for the Swedish magazine \emph{Popul\"{a}r Astronomisk Tidskrift},
none of which are
presently in ADS (There is an effort to make it so). Some of the articles
are on historical figures such as Tycho Brahe, obituaries, and 
discoveries or reports from meetings. He also wrote several popular
science books and encyclopedia articles that are not included.} 
The recording of citations in Lundmark's papers were approached in the same
manner as for Hubble.

To ensure these are relatively comparable figures ADS was used to see how often
other authors mentioned Lundmark and Hubble in their articles.
Using the ADSLabs Fulltext Service\footnote{Currently residing
at http://labs.adsabs.harvard.edu/fulltext} Statistics were compiled
on how many times authors mentioned Lundmark and Hubble by name
from 1920 through 1930.  Astronomers did not consistently include a full
citation to other authors' works, but often only referenced
the author's name.  All papers have been eliminated where Lundmark
or Hubble have cited/mentioned themselves.\footnote{There are certain to
be a number of complaints about this methodology. For example, perhaps
Lundmark simply likes to includes lots of citations, or Hubble only
cites ``big shots", etc. Those and others are certainly valid complaints,
but they should not distract one from making an attempt.}
From Appendices \ref{appendixHubble} and \ref{appendixLundmark}, it is apparent that Lundmark did
not simply fill up his papers with citations for the sake of doing so, but 
because he had a broad knowledge of the scientific literature in his field
of study. This is confirmed by \citet[][p. 130]{Holmberg1999reaching}:
\begin{quote}
He had read widely and perhaps knew the astronomical literature better
than most astronomers...
\end{quote}

\begin{figure}[ht]
\center{\plotone{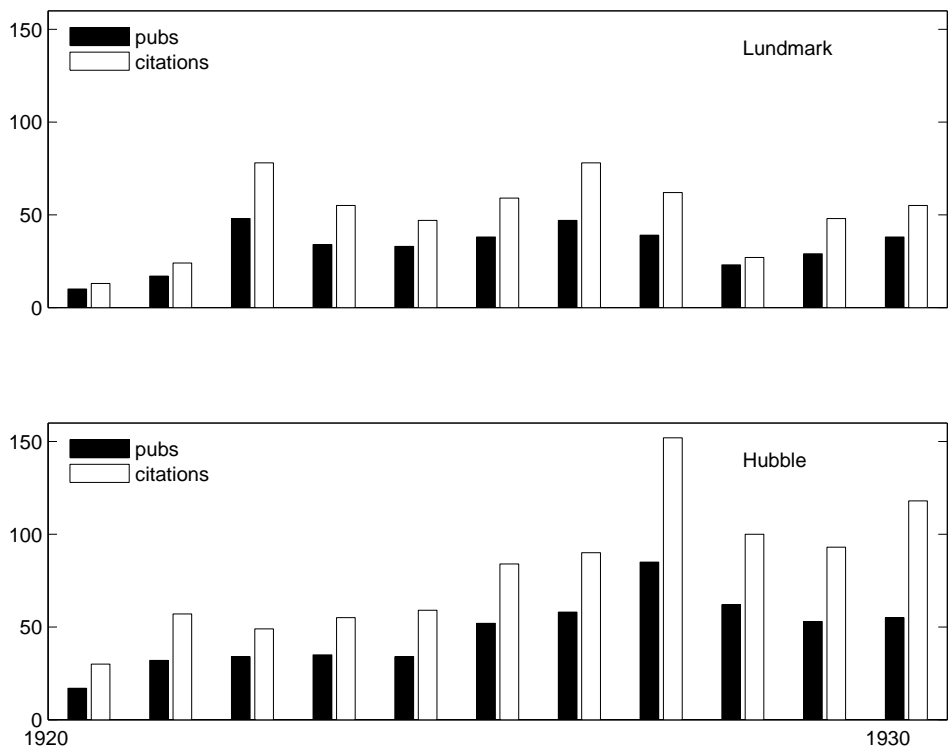}}
\caption{Top: Black are the number of publications 1920 through 1930 that cite
Knut Lundmark per year, white are the total number of citations per year.
Bottom: The same for Hubble.}\label{wayfig01}
\end{figure}

While Hubble was inconsistent in his citations, this
inconsistency was not necessarily reflected by the status
of the individual he did or did not cite. To attempt to properly quantify
these kinds of tendencies would require a much larger and sophisticated
effort than that provided herein;
nonetheless, it is clear from Figure \ref{wayfig01} that in the early half of
the 1920s it is Lundmark who is ``cited" more frequently, while in the latter half
it is Hubble.  This is not surprising given not only the facilities with which
Hubble was able to conduct his research (including the largest aperture
telescope in the world at Mt. Wilson), but also Hubble's success at promoting
himself and Mt. Wilson as described above.

\begin{figure}[ht]
\center{\plotone{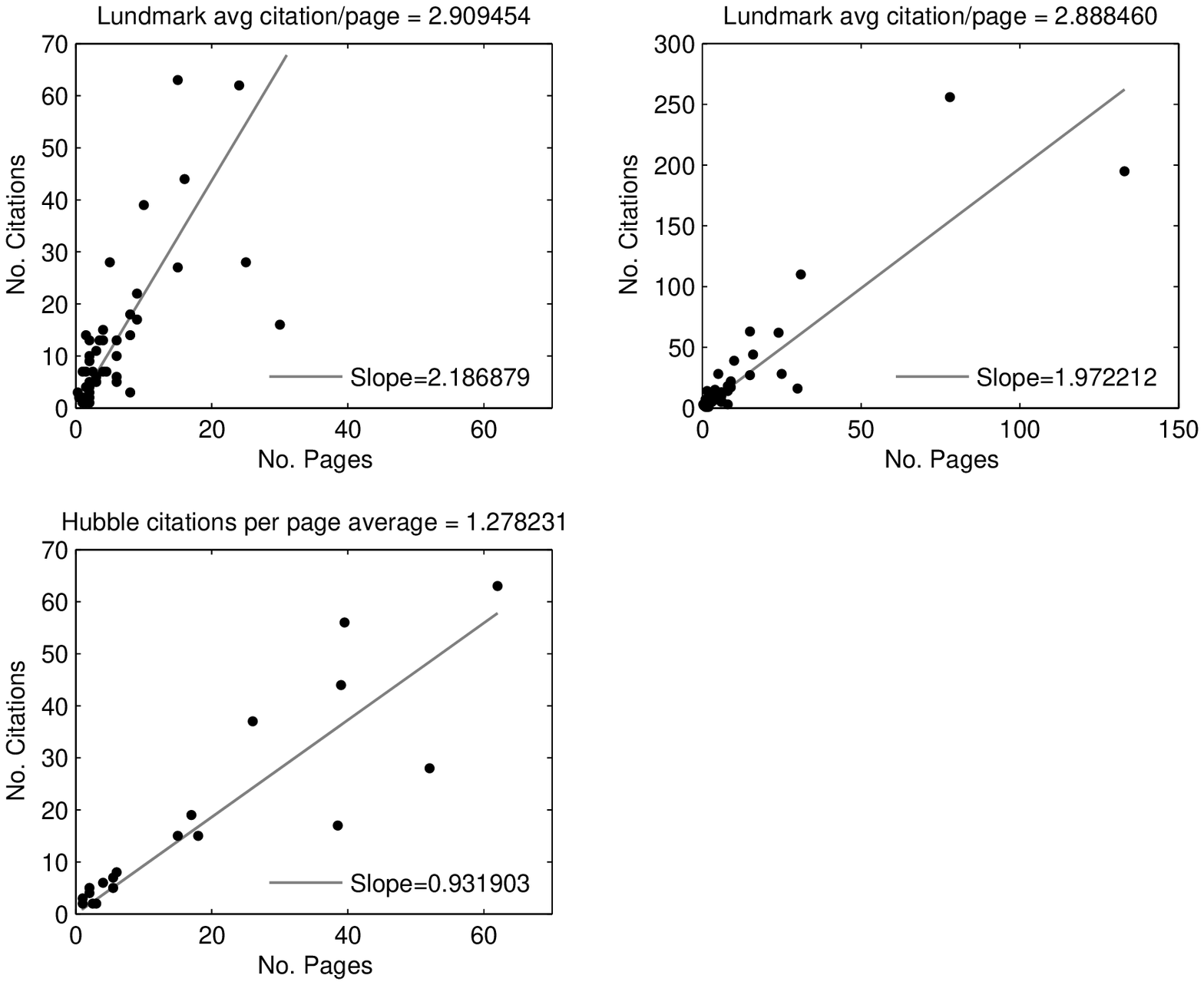}}
\caption{Top Left: Citations per page in Lundmark 1920--1930 publications of less than 70 pages.
Top right: Citation per page in all Lundmark 1920--1930 publications. Bottom left: Hubble
citations per page in all publications 1920--1930.}\label{wayfig02}
\end{figure}

Figure \ref{wayfig02} shows how often Lundmark and Hubble cited other authors.
Unfortunately there is no way to accomplish
this within ADS at present.\footnote{The author is currently working with
leading text data mining researchers to make this possible in the future.}
To obtain these numbers was difficult and required reading through every paper
and counting the number of actual citations, not just names, to
authors.\footnote{By citation we are being
flexible in that a citation can simply refer to an author's
work by name (without a journal reference), but they can only be ``cited" once
per paper (this is sometimes difficult to discern).}
Figure \ref{wayfig02} shows a clear trend in that  Lundmark cites authors
almost twice as frequently as Hubble. If one allows all
of Lundmark's publications (Hubble's longest was 65 pages, while Lundmark's
has 3 over 100 including one over 250) Lundmark is still over a factor
of two higher in citation rate per page.

\begin{figure}[ht]
\center{\plotone{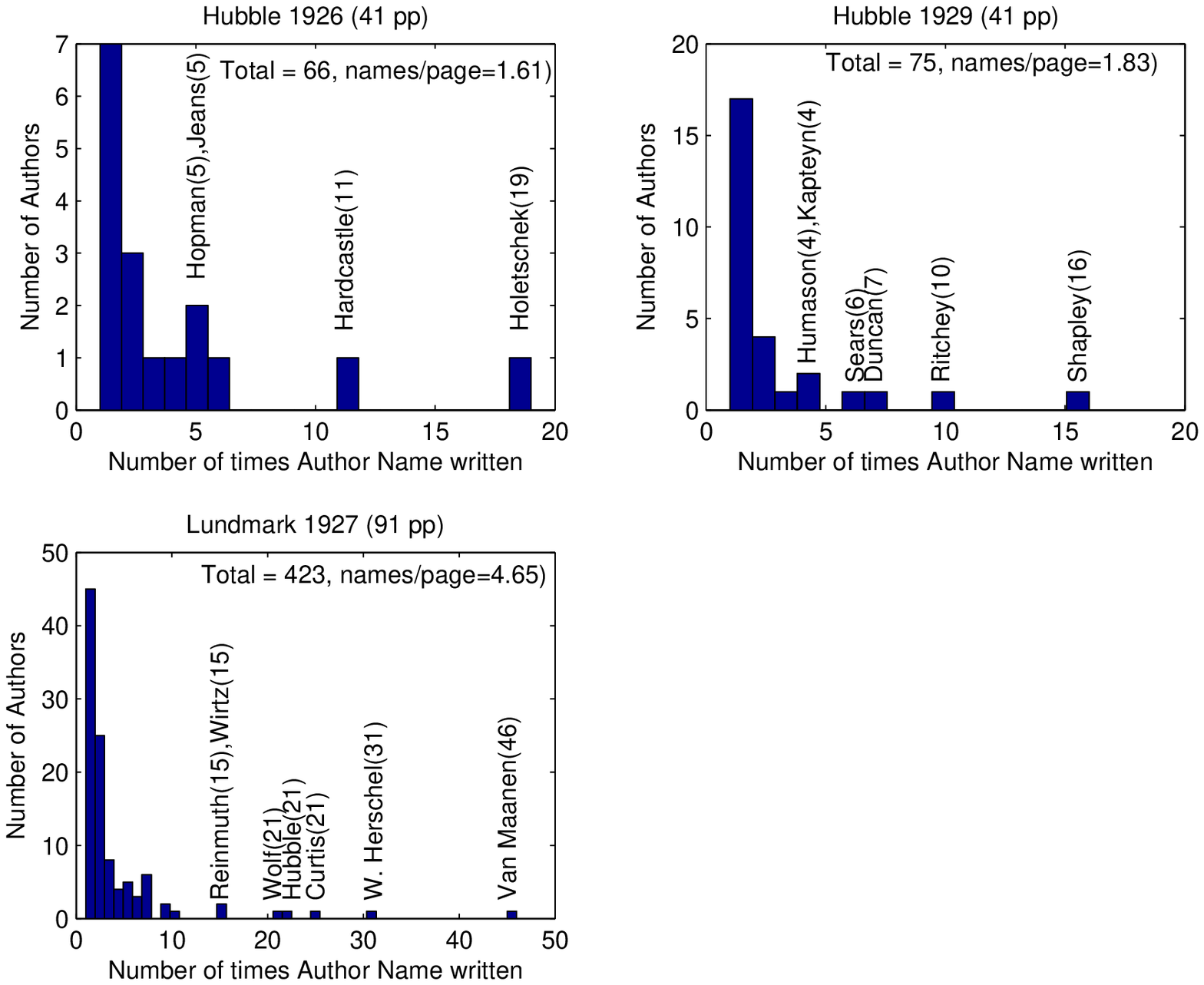}}
\caption{Top Left: Number of author names (including duplicates) found in 
\citep{Hubble1926ApJ....63..236H}. Top Right: same for \citep{Hubble1929ApJ....69..103H}.
Lower Left: Same for \citep{Lundmark1927UGC..........1L}. Note that all of the
scales on the y-axes are distinct, as are the x-axes between the
top two and bottom plot.}\label{wayfig03}
\end{figure}

A more specific comparison can be made by examining two papers by Hubble and
one by Lundmark (Figure \ref{wayfig03}).  The first one of Hubble
was 51 pages in length \citep{Hubble1926ApJ....63..236H};
not including pages with tables or photos taking up a full page it is only 41.
A second Hubble paper \citep{Hubble1929ApJ....69..103H} is 63 pages in length;
without full page tables/photos only 41. For Lundmark a paper 195 pages in
length was chosen \citep{Lundmark1927UGC..........1L} (91 eliminating
full pages with tables/photos). In both Hubble and Lundmark's
papers the number of author names per paper was counted (not including their own
names).\footnote{This is a looser criterion from that above where only
\emph{citation} related names were compiled.} In the two
Hubble papers there were a total of 66 and 75 names mentioned (many repeated of
course), which comes to 1.61 and 1.83 names per page. For Lundmark's paper there
were 423 names mentioned which comes to 4.65 names per page.

Of course the raw numbers as presented in Figure \ref{wayfig03}
should be renormalized to the number of pages, but one can see
that even dividing Lundmark's numbers by a little over a factor of
two would reveal that he still mentioned his colleagues much more frequently
than Hubble. It should also be obvious to the knowledgeable reader
that all of the more highly-cited names shown are the expected authors
in this particular domain.

One could also extend this type of study in numerous directions to better
quantify these effects. For example, it could be interesting to compare
how often authors cite others who their work explicitly relies upon.
Do the works on classification of Herschel, Reynolds, or Wolf show up
in the classification works of Lundmark or Hubble? We know from above that
\cite{Reynolds1920MNRAS..80..746R} was never found in any of Hubble's work, but
what about the others who would have influenced Hubble?

\section{The Making of Mythic Heroes}\label{sec:mythic}

It would be inappropriate to suggest that Hubble was irrelevant to the history
astronomy, but is clear that many of the advances discussed above would
have happened within a short period of time of their original
discovery even if Hubble had never worked as an astronomer.\footnote{As
a (distant) analogy lets consider the accomplishments of Alexander the Great.
Were his accomplishments inevitable because the Macedonian state was ready for a
strong ruler after the death of his father Philip?  Wasn't the Macedonian
version of the Greek phalanx with its longer spear and the high level of
training required to master this technology in place well before Alexander? If
one answers yes to these things, then perhaps another
``strong man" could have come to the
helm of the state and accomplished much the same as Alexander.
Or if Philipp had lived he surely would
have attempted many of the same things that Alexander attempted.}
This may be contrary to the assertion by \citet[][p. 72]{Smith2009JHA....40...71S}
that technological determinism alone does not explain Hubble's discoveries.
Indeed, technological determinism, the belief that such discoveries
were inevitable given facilities such as the large telescopes at Mt. Wilson,
is not needed. Instead one may consider historical determinism ``lite." 
Clearly there were many
astronomers working in these topics and steady progress was being made in each
field.  If Hubble had not found Cepheids in spiral nebulae in 1924/25, then
someone else surely would have within a few years. Lundmark's classification
scheme, or the earlier scheme of Reynolds for spirals, could have easily
replaced Hubble's at some level and the linear distance-velocity
relation had already been postulated by Lema\^{i}tre in 1927.
Without Hubble it is clear that within a short period of time someone else
would have been given credit for each of these initial discoveries,
Stigler's Law of Eponymy notwithstanding
\citep{Merton1965shoulders,Stigler1980Eponymy}.\footnote{The most famous
quote from Stigler's paper is: ``No scientific
discovery is named after its discoverer."}
Hubble's success in gaining credit for his classification scheme
and linear distance-velocity relation may be related to his verification of the
Island Universe hypothesis -- after the latter, his prominence as a major player
in astronomy was affirmed.  As pointed out by \cite{Merton1968Science}
credit for simultaneous (or nearly so) discoveries is usually given
to eminent scientists over lesser-known ones.\footnote{This has been
termed ``The Matthew Effect."}

One may also consider the competition between the Lick, Lowell and
Mt. Wilson Observatories in early observational cosmology.
Because Lowell Observatory did
not enjoy a high level of esteem, \footnote{The poor reputation was initially
due to Percival Lowell's claim to have discovered canals engineered by
intelligent beings on Mars.} it may have taken Slipher more effort than
other professional astronomers
to convince the community of the validity of his initial discoveries
(see contributions by Joseph S. Tenn, Robert Smith and others in this book).
In addition, Slipher had a modest personality and was not given to
boasting or promoting his accomplishments in public. On the other hand, consider
Lundmark or Wirtz who did not have regular access to telescopes,
instrumentation and support facilities like that of Lick (Clark 36-in [1888] and
Crossley 36-in [1905]), Mt. Wilson (60-in [1908] and Hooker 100-in [1917]),
or Yerkes (40-in [1895]).\footnote{Lundmark did
visit and observe at Mt. Wilson before his falling out with Hubble over
their classification schemes. He also spent time at Lick and Mt. Wilson over
a period of two years (1921-22) thanks to the Sweden-America
foundation \citep[][p. 94]{Holmberg1999reaching}. At that time he was able to
borrow plates from Lick Observatory. In 1929 and in 1932 he again
visited several American observatories (including Mt. Wilson) to utilize
their plate collections for his Lund General Catalogue project
(Ibid., pp. 109-116).}
Did Lundmark and Wirtz have to write more cogent and highly interesting/readable
papers to get the community to follow their research in a ``competition for attention"
\citep[][p. 480]{Collins1975conflict}? If so, then we know that Lundmark
succeeded at some level because his 1920 thesis \citep{Lundmark1920KSVH...60....1L}
was read by astronomers such as Shapley and van Maanen before he had
even arrived at Mt. Wilson and Lick Observatories
\citep[][p. 94]{Holmberg1999reaching}. In fact, Lundmark's work was so highly
regarded that a ``Memorial Volume" was edited by
Martin \cite{Johnson1961Lundmark} and published three years after
Lundmark's death. It included contributions by
a number of highly regarded astronomers such as Milton Humason (p. 26),
Harlow Shapley (p. 37), Boris Vorontsov-Velyaminov (p. 43), Fritz Zwicky (p. 55) and
Gustaf Str\"{o}mberg (p. 95).

Of course language may have also played a role. None of
Wirtz's major works were published in English,\footnote{Much of observational
cosmology research was being produced at American observatories, while
much of the theory was being promoted by Eddington, de Sitter and others in
English language publications in Europe during the trying economic times
after World War I.}
whereas Lundmark wrote mainly in English language
journals from around 1920.\footnote{Lundmark also published in Swedish in popular
science publications and in English and Swedish
in journals like \emph{Arxiv f\"{or}
Matematik, Astronomi och Fysik}, and \emph{Lund or Uppsala Observatory Reports}.}
Is it possible that an astronomer like Hubble coming from the premier observatory
didn't need to be as explicit in his citing of previous work because people
{\it had} to read and utilize his results regardless of the quality
of the background scholarship?

There is also the issue of the relative decline of research activities
at smaller institutions that took place in the U.S.A. (and Europe) around
the turn of the century given that the latest observational astronomy research
required large amounts of capital \citep[][Chapter 7]{Lankford1997american}.
As well, the First World War did not do anything positive for the facilities at
European institutions in this sense.  Simply consider Shapley's move from
Mt. Wilson to Harvard around 1925. Clearly he knew he was giving up access to one
of the finest observatories in the world to work at an institution
with inferior equipment, although as director of Harvard Observatory
he was to obtain sufficient research funding for otherwise large-scale projects.

One should take account of these changes in order to better quantify how the
attitudes of scientists working at the premier institutions could have changed
and how credit for discoveries was subsequently awarded.
While it may be a stretch to quote from William Pitt that
``Unlimited power is apt to corrupt the minds of those who possess it." One should
not rule out such effects on the minds of successful scientists.
Aggression and competition surely play some kind
of role -- a role which sociologists have already explored
in a number of scientific contexts, including the ``necessity" to
defend one's position and ideas \citep[][p. 187]{Lankford1997american}.
\citet[][p. 482]{Collins1975conflict} has also suggested that as scientists
move into higher status positions they may take on more aggressive roles.
Certainly we have seen aspects of this behavior in Hubble in his footnote
dispute with Lundmark (also see Chapter 11 in \cite{Christianson1996}),
and his persistent tendency to defend his discoveries as sole achievements
of himself and Mt. Wilson.\footnote{Hubble wrote to de Sitter
in 1930 \citep{Hubble1930deSitter} in response to de Sitter's recent
publication \citep{deSitter1930BAN.....5..157D} of a velocity-distance relation
and his lack of sufficient credit to Hubble: ``I consider the velocity-distance
relation, its formulation, testing and confirmation, as a Mount Wilson
contribution and I am deeply concerned in its recognition as such."}

It has been argued \citep[see][]{KraghSmith2003HisSc..41..141K} that much
of Hubble's fame at-large came after his death
in 1953.\footnote{Although in 1934 Lundmark was already lamenting a wave
of ``Hubbleianism" \citep[][p. 100]{Holmberg1999reaching}}
This retrospective view of Hubble's accomplishments would certainly fit in with
the well-known hypothesis of Thomas \cite{Kuhn1962structure} that:
\begin{quote}
There is a persistent tendency to make the history of science look
linear or cumulative, a tendency that even affects scientists looking back at
their own research.
\end{quote}

Of course, a similar analysis could apply to many scientific discoveries of
importance in the early 20th century.  An outstanding example is Henrietta
Levitt's discovery of a period--luminosity relation for Cepheid variable stars
and the number of scientists it took to place it on a reliable and
accurate footing.

\section{Conclusion}

Can one say anything definitive about the credit that Hubble has received
for the seminal discoveries discussed herein, and his lack of acknowledgement
of the work of others? At the present time it does not seem possible to quantify
these observations. The line of research presented in this work is only
an initial attempt. With new text data mining technologies growing in
strength one should be able to better quantify some of the assertions above
in the near future. Given what is known today it would not be fair to suggest
that Hubble was a Lavoisier-like figure who regularly claimed the discoveries of
others as his own \citep[][pp. 206-9]{Butterfield1959origins}, but that he
was inconsistent in awarding credit.

Future researchers are certain to mine the literature in detail to examine how
major scientists cited (or not) their colleagues, but will this influence the
writers of today?  Perhaps this could be accomplished by demonstrating that over
the long-term one may be discredited for neglecting to cite relevant work
that one relied upon.

Take two recent books from the 2000s that were highly read in the
scientific community: Stephen Wolfram's book \citep{Wolfram2002new}
and that of Roger Penrose \citep{Penrose2005road}. Wolfram's book contains
almost no citations to other work, while Penrose makes a valiant attempt to cite
others for a book even broader in scope than Wolfram's. Needless to say Wolfram
was pilloried in the scientific and popular press for this lack of attribution
and general belief that his ideas have broader application than
possible \citep[e.g.][]{Casti2002Science,AmScientist2002Wolfram,Economist2002Wolfram}.
Perhaps the general public agrees as well: Wolfram has received 3 stars out
of 5 (from 344 reviews) on Amazon.com as of 2012/12/05 (with a large
number of 1 stars (102) and 2 stars (61). Penrose received 4 out of 5
(from 204 reviews) with very few 1 (12) or 2 (12) stars.
Penrose's bibliography is 30 pages in length (pp. 1050--1080) and has
received rather more favorable reviews
\citep[e.g.][]{Johnson2005NYT.BookRev,Blank2006AMS.BookRev}.

How credit is awarded for a discovery is often a complex issue and should not
be oversimplified -- yet this happens time and again. Another well-known example
in this field is the discovery of the Cosmic Microwave Background
(see \cite{AlpherHerman1988PT,Gribbin1998search,Kragh1999cosmology}).

The problem is larger than awarding credit within a given field
as outsiders pick anecdotal stories from astronomy, and not only get
them wrong, but oversimplify them.\footnote{A recent example would
be \cite{taleb2010black} in his \emph{Inadvertent Discoveries} subsection of
Chapter 11 (on page 168 of the paperback edition) where he discussed the
discovery of the Cosmic Microwave Background Radiation.}
This can also happen in the case of professional astronomers/physicists
\citep[see][]{Greene2011NYT.OpED}.\footnote{To take one of many
mis-statements in this piece: ``In 1929, the American astronomer Edwin Hubble 
discovered that distant galaxies are all rushing away from us."}
One only needs to read a smattering of the contributions in this
book to understand how misguided Greene was in his assumptions.
Perhaps, as a community, astronomers can learn to do better and
this book could be the beginning, at least in this particular domain.

\acknowledgements 
This research has made use of NASA's Astrophysics Data System Bibliographic
Services.

Thanks are due to the ever-efficient librarians at The Goddard Institute for
Space Studies, Zoe Wai and Josefina Mora, without whom many of the references
in this paper would have taken much more time to locate. I also wish to thank
Samuel Regandell, Bengt Edvardsson and Kjell Lundgren of the Astronomy department
at Uppsala University for their help in locating (and scanning) some
hard-to-obtain manuscripts by Knut Lundmark in their basement storage. My wife
Elisabeth also deserves praise for patiently listening to my many complaints
about the literature and soothing my stress from co-organizing this conference
and Proceedings.

This work has also benefited from comments by Bengt Edvardsson, John Peacock,
and Jeffrey Scargle and my excellent co-Editor Deidre Hunter.

\newpage

\appendix
\section{Hubble's published papers 1 Jan. 1920 through 31 Dec. 1930}\label{appendixHubble}
\begin{table}[ht!]
\begin{center}
\small
\begin{tabular}{|c|l|c|c|l|} \hline
No.& Reference & Pages&  Citations & Notes\tablenotemark{a} \\
\hline
1  &  \cite{Hubble1920CMWCI.187....1S} & 15.0 & 15 & + \\
2  &  \cite{Hubble1920PYerO...4....2H} & 18.0 & 15 & + \\
3  &  \cite{Hubble1920ApJ....52....8S} & --   & -- & See ref. 1 \\
4  &  \cite{Hubble1921PASP...33..174H} & --   & -- & N/A not Hubble's work?\\
5  &  \cite{Hubble1922CMWCI.241....1H} & 39.0 & 44 & + \\
6  &  \cite{Hubble1922CMWCI.250....1H} & 38.5 & 17 & + \\
7  &  \cite{Hubble1922ApJ....56..162H} & --   & -- & See ref. 5 \\
8  &  \cite{Hubble1922PASP...34..292H} &  2.0 &  4 & + \\
9  &  \cite{Hubble1922ApJ....56..400H} & --   & -- & See ref. 6\\
10 &  \cite{Hubble1923PA.....31..644H} &  1.0 &  3 & * \\
11 &  \cite{Hubble1923PASP...35..261H} &  2.5 &  2 & --\\
12 &  \cite{Hubble1925CMWCI.304....1H} & 26.0 & 37 & ! \\
13 &  \cite{Hubble1925PA.....33..252H} &  3.0 &  2 & --\\
14 &  \cite{Hubble1925Obs....48..139H} & --   & -- & See ref. 13\\
15 &  \cite{Hubble1925ApJ....62..409H} & --   & -- & See ref. 12\\
16 &  \cite{Hubble1926CMWCI.310....1H} & 39.5 & 56 & + \\
17 &  \cite{Hubble1926ApJ....64..321H} & --   & -- & See ref. 17\\
18 &  \cite{Hubble1926CMWCI.324....1H} & 52.0 & 28 & --\\
19 &  \cite{Hubble1926ApJ....63..236H} & --   & -- & See ref. 16\\
20 &  \cite{Hubble1926PASP...38..258H} &  2.0 &  5 & * \\
21 &  \cite{Hubble1927ASPL....1...35H} & --   & -- & N/A Popular Article\\
22 &  \cite{Hubble1927CMWCI.335....1H} &  5.5 &  7 & --\\
23 &  \cite{Hubble1927CoMtW...3...23H} & --   & -- & See ref. 34\\
24 &  \cite{Hubble1927CoMtW...3...85H} & --   & -- & N/A written in 1934?\\
25 &  \cite{Hubble1927CoMtW...3..111H} & --   & -- & N/A written in 1936?\\
26 &  \cite{Hubble1927PAAS....5...63H} & --   & -- & See ref. 10\\
27 &  \cite{Hubble1927PAAS....5..261H} &  3.0 &  2 & \\
28 &  \cite{Hubble1927ApJ....66...59H} & --   & -- & See ref. 22\\
29 &  \cite{Hubble1927Obs....50..276H} &  5.5 &  5 & + \\
30 &  \cite{Hubble1928ASPL....1...55H} &  4.0 &  6 & * \\
31 &  \cite{Hubble1929ASPL....1...93H} & --   & -- & N/A Article ``excerpt"\\
32 &  \cite{Hubble1929CMWCI.376....1H} & --   & -- & See ref. 33\\
33 &  \cite{Hubble1929ApJ....69..103H} & 62.0 & 63 & ! \\
34 &  \cite{Hubble1929PNAS...15..168H} &  6.0 &  8 & + \\
35 &  \cite{Hubble1930CMWCI.398....1H} & --   & -- & See ref. 37\\
36 &  \cite{Hubble1930PA.....38R.598H} &  1.0 &  2 & + \\
37 &  \cite{Hubble1930ApJ....71..231H} & 17.0 & 19 & + \\
\hline
\end{tabular}
\tablenotetext{a}{* No journal/book citations found, only mentions names.\\
+ At least one``reference" but without any citation, otherwise one or more normal citations.\\
! Includes +/* and/or reference to list of coordinates.\\
? Cannot obtain this reference.}
\end{center}
\end{table}

\newpage

\section{Lundmark's published papers 1 Jan. 1920 through 31 Dec. 1930}\label{appendixLundmark}
\begin{table}[ht!]
\begin{center}
\small
\begin{tabular}{|c|l|c|c|l|} \hline
No.& Reference & Pages&  Citations & Notes\tablenotemark{a} \\
\hline
1  &  \cite{Lundmark1920KSVH...60....1L} & 78.0 &256 & --\\
2  &  \cite{Lundmark1921AN....213...93L} &  2.0 & 10 & --\\
3  &  \cite{Lundmark1921AN....213..315L} &  2.0 & 10 & --\\
4  &  \cite{Lundmark1921PASP...33..219L} &  1.5 &  1 & --\\
5  &  \cite{Lundmark1921PASP...33..225L} & 15.0 & 27 & ! \\
6  &  \cite{Lundmark1921PASP...33..271L} &  1.5 &  7 & * \\
7  &  \cite{Lundmark1921PASP...33..314L} &  3.0 & 11 & ! \\
8  &  \cite{Lundmark1921PASP...33..316L} &  3.0 &  5 & * \\
9  &  \cite{Lundmark1921PASP...33..324L} &  4.0 & 15 & ! \\
10 &  \cite{Lundmark1922LicOB..10..149L} &  3.5 & 13 & --\\
11 &  \cite{Lundmark1922LicOB..10..153L} &  4.0 & 13 & + \\
12 &  \cite{Lundmark1922PAAS....4..368L} &  2.0 &  1 & * \\
13 &  \cite{Lundmark1922PAAS....4..370L} &  1.0 &  2 & * \\
14 &  \cite{Lundmark1922PAAS....4Q.371L} &  1.0 &  7 & * \\
15 &  \cite{Lundmark1922PASP...34...40L} & 10.0 & 39 & ! \\
16 &  \cite{Lundmark1922PASP...34...53L} &  0.5 &  2 & * \\
17 &  \cite{Lundmark1922PASP...34..108L} &  8.0 & 18 & +!\\
18 &  \cite{Lundmark1922PASP...34..126L} &  2.0 & 13 & +!\\
19 &  \cite{Lundmark1922MNRAS..82..495L} & 16.0 & 44 & --\\
20 &  \cite{Lundmark1922PASP...34..147L} &  9.0 & 22 & + \\
21 &  \cite{Lundmark1922PASP...34..191L} &  9.0 & 17 & + \\
22 &  \cite{Lundmark1922PASP...34..207L} &  5.0 & 28 & + \\
23 &  \cite{Lundmark1922PASP...34..225L} &  1.5 & 14 & *+!\\
24 &  \cite{Lundmark1922PASP...34..229L} &  0.3 &  3 & --\\
25 &  \cite{Lundmark1922PASP...34..292H} &  2.0 &  4 & + \\
26 &  \cite{Lundmark1922PASP...34..357L} &  8.0 &  3 & --\\
27 &  \cite{Lundmark1923PA.....31..239L} &  2.0 &  2 & --\\
28 &  \cite{Lundmark1923PA.....31..241L} &  2.0 &  9 & +!\\
29 &  \cite{Lundmark1923PA.....31..455L} &  3.0 &  6 & +!\\
30 &  \cite{Lundmark1923PASP...35...95L} & 25.0 & 28 & --\\
31 &  \cite{Lundmark1923MNRAS..83..470L} &  4.5 &  7 & + \\
32 &  \cite{Lundmark1923AJ.....35...93L} &  2.5 &  5 & + \\
33 &  \cite{Lundmark1923PASP...35..318L} &  2.0 &  4 & + \\
34 &  \cite{Lundmark1924MNRAS..84..747L} & 24.0 & 62 & + \\
35 &  \cite{Lundmark1924Obs....47..276L} &  4.0 &  7 & + \\
36 &  \cite{Lundmark1924Obs....47..279L} &  2.5 &  7 & + \\
37 &  \cite{Lundmark1925MNRAS..85..865L} & 31.0 &110 & +!\\
38 &  \cite{Lundmark1926CMWCI.308....1L} &  6.0 &  6 & + \\
39 &  \cite{Lundmark1926VdAG...61..254L} & --   & -- & ? \\
40 &  \cite{Lundmark1926ApJ....63...67L} & --   & -- & See ref. 38\\
41 &  \cite{Lundmark1926ArMAF..19B...8L} &  2.0 &  5 & * \\
42 &  \cite{Lundmark1926ArMAF..19B...9L} &  6.0 & 13 & + \\
43 &  \cite{Lundmark1927ArMAF..20B...3L} &  6.0 &  5 & --\\
44 &  \cite{Lundmark1927ArMAF..20A..12L} & --   & -- & ?\\
45 &  \cite{Lundmark1927ArMAF..20A..13L} & --   & -- & ?\\
\hline
\end{tabular}
\end{center}
\end{table}

\begin{table}[ht!]
\begin{center}
\small
\begin{tabular}{|c|l|c|c|l|} \hline
No.& Reference & Pages&  Citations & Notes\tablenotemark{a} \\
\hline
46 &  \cite{Lundmark1927ArMAF..20A..18L} & --   & -- & ?\\
47 &  \cite{Lundmark1927ArMAF..20A..20L} & 15.0 & 63 & + \\
48 &  \cite{Lundmark1927PAAS....5...16L} &  2.0 &  3 & + \\
49 &  \cite{Lundmark1927PAAS....5...18L} &  2.0 &  5 & + \\
50 &  \cite{Lundmark1927UGC..........1L} &133.0 &195 & ! \\
51 &  \cite{Lundmark1928ArMAF..21A...8L} & --   & -- & ?\\
52 &  \cite{Lundmark1928ArMAF..21A...9L} &  6.0 & 10 & + \\
53 &  \cite{Lundmark1928ArMAF..21A..10L} &  8.0 & 14 & + \\
54 &  \cite{Lundmark1928MfAOU...41....L} & --   & -- & ? \\
55 &  \cite{Lundmark1930PA.....38...26L} &  1.5 &  4 & * \\
56 &  \cite{Lundmark1930PA.....38...27L} &  1.0 &  1 & * \\
57 &  \cite{Lundmark1930PASP...42...23L} & 30.0 & 16 & * \\
58 &  \cite{Lundmark1930PASP...42...31L} &  3.0 &  5 & * \\
\hline
\end{tabular}
\tablenotetext{a}{* No journal/book citations found, only mentions names.\\
+ At least one``reference" but without any citation, otherwise one or more normal citations.\\
! Includes +/* and/or reference to list of coordinates.\\
? Could not obtain this reference.}
\end{center}
\end{table}

\newpage

\section{Lundmark's reply to Hubble May 1927}\label{appendixlundmarkreply}
Lundmark's reply to Hubble's accusation of plagiarism was presented
as a paper to the Royal Society of Science of Upsala on May 6, 1927
and later published in the \emph{New Proceedings of the Royal Society of
Science of Upsala} \footnote{Nova Acta Regiae Societatis Scientiarum
Upsaliensis, Volumen Extra Ordinem Editum 1927} and provides a unique
look at how at least one of Hubble's peers viewed him at that time.
I produce this quote (verbatim) here since this volume was, until recently,
very difficult to obtain.

\begin{quote}
``In his paper, Extragalactic nebulae, Aph. J. 64:321, 1926, E. P. Hubble makes
an attack on me which is written in such a tone that I hesitate to give any
answer at all. Still, I may take the occasion to state a few facts.

I was present at the Cambridge meeting of the Astronomical Union.

I was not then a member of the Commission of Nebulae.

I did not have any, access whatsoever to the memorandum or to other writings of
E. P. Hubble, neither did I have access to the report of nebulae (which does not
give details of Hubble's classification) until at the end of the meeting, neither
did I recognize until I obtained a letter from Hubble at the end of 1926 that he
had made another classification of nebulae than the one published in his paper,
A general study of the Diffuse Galactic Nebulae, Mt Wils. Contr. No. 241, 1922.

As much as I heard of the discussion in the committee of nebulae the only
question was if the terms $\gg$galactic$\gg$ and $\gg$extragalactic nebulae$\gg$ should be
accepted, From the discussion I got the impression that the intention of Hubble
was to force through his nomenclature, One of the members told me outside
the discussion that Hubble had suggested the
subdivision $\gg$logarithmic spirals$\gg$ but
I did not understand that this suggestion was given in any memorandum to the
Union, Now when reading Hubble's paper I am glad to note that he seems not to
have carried out the unhappy idea introducing the term $\gg$logarithmic spirals$\gg$
Slight changes in his classification might have been introduced since the
Cambridge meeting.

Hubble's statement that my classification except for nomenclature is practically
identical with the one submitted by him is {\it not correct}. Hubble classifies
his subgroups according to eccentricity or form of the spirals or degree of
development while I use the degree of concentration towards the centre. As to
the three main groups, elliptical, spiral and magellanic nebulae it may be of
interest to note that the two first are slightly older than Hubble and myself.
The term elliptical nebulae thus is used by Alexander in 1852 and the term spiral
by Rosse in 1845; The importance of the magellanic group has been pointed out by
myself Observatory 47, 277, 1924 earlier than by Hubble. \emph{As to Hubble's way
of acknowledging his predecessors I have no reason to enter upon this
question here.}" (the latter is our emphasis).
\end{quote}

\nocite{Duerbeck2001mkt..book..231D}
\nocite{Hertzsprung1913AN....196..201H,Fox1913Sci....37..639F}
\nocite{Block.Freeman2008book}
\nocite{Leavitt1912HarCi.173....1L}

\bibliography{way}

\end{document}